\pdfoutput=1
\documentclass[final,3p,times,twocolumn]{elsarticle}
 \biboptions{comma,sort&compress}
\usepackage{here}
 \usepackage{graphicx}
  \usepackage{epsfig}

 \usepackage{fancyhdr}                   
\newcommand {\beq} {\begin{equation}}
\newcommand {\eeq} {\end{equation}}
  \newcommand {\ber}{\begin{eqnarray*}}
  \newcommand {\eer} {\end{eqnarray*}}
\newcommand {\beqn}{\begin{eqnarray}}
  \newcommand {\eeqn} {\end{eqnarray}}

\newcommand{\Dslash}{\,{\raise.15ex\hbox{/}\mkern-12mu D}}

\newcommand{\gsim}{\lower.7ex\hbox{$
\;\stackrel{\textstyle>}{\sim}\;$}}
\newcommand{\lsim}{\lower.7ex\hbox{$
\;\stackrel{\textstyle<}{\sim}\;$}}

\journal{Nuc. Phys. (Proc. Suppl.)}

\begin{document}

\begin{frontmatter}

\title{ 
\begin{flushright}
{\small FTPI-MINN-10/28, 
UMN-TH-2921/10\\
October 7, 2010}
\end{flushright}
 QCD Sum Rules: Bridging the Gap  
   between Short and Large Distances}

\author{ M. Shifman }

 \address { William I. Fine Theoretical Physics Institute, University of Minnesota,
  Minneapolis, MN 55455, USA$^*$\corref{cor1}\\
 and\\
 Jefferson Physical Laboratory, Harvard University, Cambridge, MA 02138, USA}
\cortext[cor1]{Permanent address}

\begin{abstract}
I discuss  aspects of the QCD sum rule method
which attracted theorists' attention in earnest at a relatively late stage and are not yet
fully solved. At first I briefly review such general
topics as the structure of the operator product expansion in QCD and
intrinsic  limitations of the quark-hadron duality concept.
In the second part I comment on holographic constructions --- a focus of the current efforts
to say something new on QCD at strong coupling. Of particular interest to me is the recent derivation
of the vacuum magnetic susceptibility due to Son and Yamamoto. Remarkably, their result is the same as that obtained previously by Vainshtein in the field-theoretic framework. For reasons which I do not understand at the moment, the Vainshtein formula, unexpectedly,
is not successfull
numerically.

This is a slightly modified version of the talk delivered at
15th International QCD Conference ``QCD 10,"
June 28 - July 3, 2010, Montpellier, France. Published in
Nuclear Physics B (Proc. Suppl.) {\bf 207--208},  298--305  (2010).

\end{abstract}

\end{frontmatter}


\section{Introduction}

I was asked to open this special session devoted to the QCD sum rules 
(sometimes referred to as the SVZ sum rules) with  a brief outline of the current status of this method
in various applications to hadronic physics and, perhaps, some historical remarks. 
Instead, I decided to do something else. 
I will completely skip the second part since quite recently I published a paper \cite{unpubintro}
where the reader can find both the description of the evolution of the 
method and relevant anecdotes (see also \cite{persistent}). I will not dwell on 
various (quite fruitful) recent applications which will be (hopefully)
covered  by other speakers.
Instead, I will focus on some ``afterthoughts" of a
general nature, some issues 
which were not explored (or not fully explored) in due time, when the SVZ sum rules were in the making.
They came into the limelight in the last ten years or so. Three main topics are
\begin{itemize}
\item
The structure of the operator product expansion (OPE) in QCD;
\item
The limitations of the quark-hadron duality;
\end{itemize}
and, finally,
\begin{itemize}
\item
Fast-forward in the past (some remarks on AdS/QCD).
\end{itemize} 

Let me first recall that the basic concept of the SVZ sum rules, its foundation, is
the following idea:

\vspace{2mm}

 The  QCD vacuum structure is complicated and not yet fully 
 analytically understood despite significant progress, especially after the advent of supersymmetry in this range of questions \cite{SW}.
 For limited purposes one can try to represent the  QCD vacuum by a few vacuum 
 expectation values (VEVs) of local operators intended to theoretically approximate various correlation functions
in an intermediate domain of distances --- between short and asymptotically large.
 The gluon and quark condensates are   most important, but one is welcome to add a few others.
 There are many reasons not to add too many, though. One of them, as I will discuss later,
 is the fact that OPE in QCD presents an asymptotic expansion. This is in contradistinction with OPE in a number of conformal field theories (with no intrinsic scale)
 in which it is believed to be better convergent or just convergent
 in the coordinate space. (Is it? A good question for a serious reflection ...).
 
\begin{figure}
\begin{center}
\includegraphics[width=3.1in]{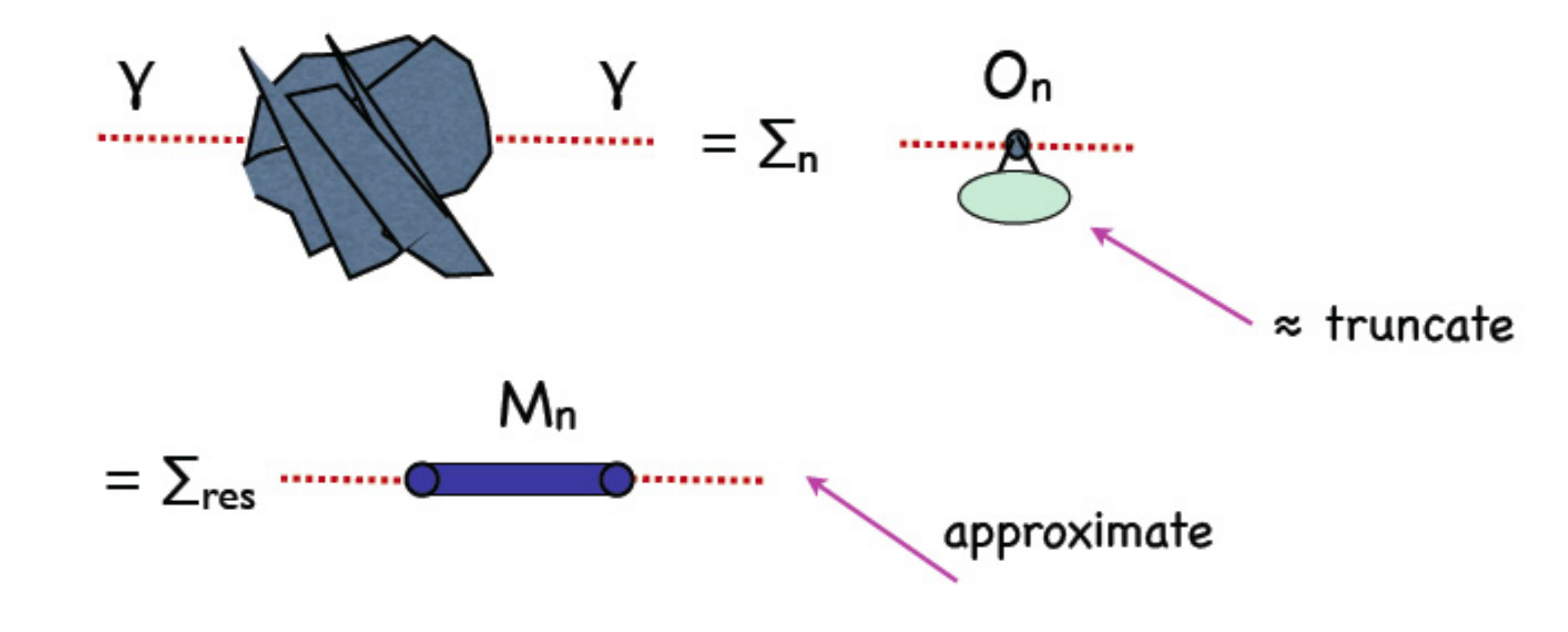}
\end{center}
\caption{\small Graphic representation of the SVZ sum rules in the correlation function
$\langle\bar\psi\gamma^\mu\psi (x)\,, \bar\psi\gamma^\nu\psi (0) \rangle$.}
\label{1}
\end{figure}

 The above condensate expansion in the intermediate domain of distances can be matched by
 the sum over hadronic states, as in Fig.~\ref{1}.
The sum is infinite, but one can enhance the contribution from the lowest-lying states
through Borelization.

\section{Operator Product Expansion}
\label{ope}

The theoretical basis of any calculation within the SVZ method is 
the 
operator product expansion. It allows one to consistently define 
and build the (truncated) condensate series for any amplitude of 
interest in the Euclidean domain. The physical picture lying behind 
OPE is a consistent separation of short 
and large  distance contributions. The latter are then represented 
by the vacuum condensates while the former are accounted for in the 
coefficient functions.  
Somewhat symbolically, the Fourier-transformed correlation functions
can be represented as
\begin{equation}
D(q^2)  =\sum_n C_n(q;\mu )\langle {\cal O}_n(\mu )\rangle 
\label{tee}
\end{equation}
where $D(q^2)$ is, say,  a two-point function,
and the normalization point $\mu$ is indicated explicitly.  
 The sum in Eq. (\ref{tee}) runs over the
Lorentz and gauge invariant  operators built from the gluon and 
quark fields.  The operator  of  the lowest (zero) dimension is 
the unit operator {\bf I},
followed by the gluon condensate $G_{\mu\nu}^2$, of dimension 
four. The four-quark condensate gives an example of dimension-six
operators. The OPE in (\ref{tee}) is, in a sense, a book-keeping procedure.

In this form Wilson designed it for theories with a 
UV fixed point at $\alpha\neq 0$ \cite{Wilson} 
(Wilson considered OPE in the coordinate space. The same was done by
 Polyakov in unpublished lectures on this topic
that circulated about this time. In this case all $x$ dependence is encoded
in the set of coefficients $C_n$.) 
Theories with the UV fixed point at $\alpha\neq 0$ are conformal at short distances,
with the power-like approach to conformality. In this case it is most natural {\em not} to introduce 
a special normalization point $\mu$. Instead, 
it is implicitly assumed to coincide with 
the external momentum $q$ or distance $x$.
OPE was thought of as the expansion in singularities
at $x\to 0$ or $q\to\infty$.

The OPE (fusion rules) in the above form are fitted for 
the conformal theories (Polyakov et.al.) where dynamics at all scales is the same.
In two dimensions the fusion rules are powerful enough to fully solve some CFTs \cite{Belavin:1984vu}.
In four-dimensional ${\mathcal N}=4$ super-Yang--Mills theories
conformal symmetry leads to miracles.

\vspace{1mm}

Alas ... our world is far less  perfect.

\vspace{2mm}

$\bullet$ {\hspace{0.3cm} In QCD: 

\vspace{1mm}

1) The UV fixed point is at $\alpha_s=0$. Approach 
to the asymptotic limit is very slow --- only logarithmic. 
Most operators cannot be defined as absolutely local. 
Anomalous dimensions are also logarithmic.

2) A dynamical scale of distances $\Lambda^{-1}$ is generated
through dimensional transmutation. Interactions on the opposite 
sides of $\Lambda^{-1}$ are drastically different.  Near 
and below $\Lambda^{-1}$ perturbative calculations are inapplicable.

3) Without introduction of a normalization point $\mu$,
a sliding boundary between the two domains --- short vs. large distances ---
construction of OPE is meaningless.

\vspace{1mm}

As a result, the Wilson--Polyakov original formula was not implemented 
as far as subleading power corrections are concerned. 
An appropriate modification of OPE was needed, fit for a very special QCD environment,
with its logarithmic approach to the asymptotic limit.
At the same time, the general Wilson's renormalization-group ideas are perfect! 
We did not change them in a conceptual way. Rather, we engineered its implementation
most appropriate for QCD. 
Wilson's idea
was finally fully adapted to the QCD environment in  \cite{NSVZ2}, with quite a number of successful 
forays beginning from 1978, in various SVZ papers.

\vspace{2mm}

Thus, Wilson's
OPE was redesigned, with an important technical
addition: we understood that power-suppressed terms of OPE need not be necessary discarded
compared to logarithmically-suppressed terms, even if one cannot sum up the
entire logarithmic series. 
They need not be smaller. I remember that implanting this idea in the minds of more formal theorists
was a difficult task. Frankly speaking, even now a formal theoretical justification for this procedure
is lacking. But it works!

\vspace{1mm}

$\bullet$ {\hspace{0.3cm} It was realized that
the degree of locality of the operators $O_n(0)$ is regulated
by an external parameter $\mu$. Even if 
the fusion operators $A$ and $B$ have vanishing anomalous dimensions
(for instance, conserved vector currents), the
coefficients $C^{AB}_n(x)$ and $O_n(0)$ depend on $\mu$, in particular, through logarithms of $\mu$, although the overall combination
$$
\sum_n C^{AB}_n(x) O_n(0)
$$
is $\mu$ independent! In practice, to make the SVZ method work for
 low-lying hadronic states one should be able to sail between Scylla and Charybdis
of contradictory requirements on $\mu$. If $\mu$ is too high or two low, one looses the predictive power!
The value of $\mu$ is to be carefully balanced. I won't go into details here, referring to \cite{snap},
and asserting that the necessary balance can be achieved in the so-called practical version of OPE.
This is a fortuitous special feature of QCD which is not necessarily shared
by simpler theories used to model various aspects of QCD.
An example of such a simple theoretical laboratory is given by two-dimensional 
$CP(N-1)$ model \cite{nsvzsigma}. Unlike the situation in $CP(N-1)$, in QCD, if
 $\mu$ is judiciously chosen, the coefficients $C^{AB}_n(0)$  are {\em mostly} determined
 by perturbation theory while the condensates $\langle O_n(0)\rangle$ are 
 {\em mostly} mostly nonperturbative. We are lucky.

\vspace{1mm}

$\bullet$ {\hspace{0.3cm} The OPE expansion runs  in powers of $1/Q^2$ and $\ln Q^2$.

$\bullet$ {\hspace{0.3cm} The OPE expansion is asymptotic  at best. 
The fact that the condensate series is factorially divergent in
high orders is rather obvious from the analytic structure of 
the polarization operator $D(Q^2)$. In a nut shell, since the cut
in $D(Q^2)$ runs all the way to infinity along the positive real 
semi-axis of $q^2$, the $1/Q^2$ expansion cannot be convergent. 
The actual argument is somewhat more  subtle
\cite{Shif1} but the final conclusion
is perfectly transparent. It is intuitively clear that the high-order tail
of the (divergent) power series gives rise to 
exponentially small corrections  $\sim\exp (-Q^\sigma )$ (where 
$\sigma$ is some critical index) in the Euclidean domain which 
oscillate and suppressed by powers of energy (or other appropriate parameters)
 in the Minkowski domain. 

The numerical value of  $\sigma$ is correlated with the rate of  
divergence of high orders
in the power series. 
This is explained in great detail in Section 8 of \cite{QHD}.
At the moment very little is known about this 
rate from first principles, if at all. The best we can do is to rely on 
toy  models \cite{QHD}.
The simplest  example is  
provided by instantons. One has to fix the
size of the instanton $\rho$ by hand,  $\rho = \rho_0$. Then the
fixed-size instanton  
contribution in the Euclidean domain is indeed exponential,  ${\cal O}(\exp (-Q\rho_0 
))$. The exponential factor is the price we pay for
transmitting the large momentum $Q$ through a soft field 
configuration
whose characteristic frequencies are or the order of $\rho_0^{-1}$. 
Being analytically continued in the Minkowski domain, the
imaginary part of the instanton contribution on the cut oscillates and is only power suppressed.

A very similar situation takes place in an alternative construction, 
the so-called resonance model worked out in Ref.~\cite{QHD}.
Both models 
 are on the market for quite a time, but --- alas ---
there were essentially no recent advances in this direction.
Any fresh ideas on possible better or more compelling  
models that would lie closer to first principles are most welcome!

Now, let me recall that not only the condensate series, but the $\alpha_s$ series 
{\em per se} are asymptotic.
Some of the high-order divergences in the $\alpha_s$ series can be absorbed in the condensates
normalized at $\mu$ (e.g. infrared  renormalons).
Others  still can show up as nonperturbative terms in the expansion coefficients
$C^{AB}_n(0)$, which, unfortunately, continue to be rather poorly controllable till present.
Our knowledge of these terms is semi-empiric. They are known to be numerically small
and neglectable
in a variety of channels (but not in all! see \cite{alike} where exceptional channels are discussed at length) under a judicial choice of the
normalization point $\mu$.
 
Concluding this part of the talk 
I can summarize the general achievements in understanding OPE in QCD as follows:
 It became clear that the overall structure of this expansion is highly complicated -- 
 much more complicated  than, say,  in the (exactly solved part of)  ${\mathcal N}=2$ 
 super-Yang--Mills \cite{SW}. Even in the Euclidean domain
 this expansion is asymptotic in various ways and includes contributions
 of different nature. It is quite timely to  try to categorize 
 those contributions
 which go beyond the truncated condensate series containing just a few terms. 
 This is difficult. Is it doable?

  \section{
Quark-hadron duality}

This section could have been entitled
``Parametrizing ignorance." If theoretical calculation is hard enough in the Euclidean domain,
the problem is immensely exacerbated upon transition
to the Minkowski kinematics which is necessary for two related reasons:
(a) estimate of the continuum in the SVZ sum rules (not so crucial
in the majority of instances); (b) predictions for highly excited states
(crucial in a number of problems, such as restoration/nonrestoration
of chiral symmetry in high excitations \cite{svregge}, hadronic widths of $\tau$ and similar, 
Regge trajectories at large $N$
and so on).
   
   If one could calculate $D(Q^2)$ in the Euclidean domain {\em exactly},
one could analytically continue the result to the Minkowski domain,
and then take the imaginary part.  The spectral density
$\rho (s)_{\rm theor}$ obtained in this way would present the {\em  exact}
theoretical prediction for the measurable hadronic  cross section.
{\em There would be no need for duality.}

In practice, our calculation of $D(Q^2)$ is approximate, for many reasons.
First, nobody is able to calculate the infinite
$\alpha_s (Q^2)$ series for the coefficient functions, let alone the infinite
condensate series. Both have to be truncated at some finite order.
A few lowest-dimension condensates that can be captured,
are known approximately.  The best we can do is
analytically continue the {\em truncated} theoretical 
expression, term by term,
from positive to negative $Q^2$. For each term in the 
expansion the imaginary
part at  positive $q^2$ (negative $Q^2$) is well-defined.  We assemble 
them together and declare
the corresponding $\rho (s)_{\rm theor}$ to be dual to
the hadronic cross section $\rho (s)_{\rm exp}$.
In the given context ``dual" means  equal.

Let me elucidate this point in more detail.
Assume that $D (Q^2)$ is calculated through $\alpha_s^2$
and $1/Q^4$, while the 
terms  $\alpha_s^3$ and $1/Q^6$ (with possible logarithms)
are dropped.
Then the theoretical quark-gluon 
spectral density, obtained as described above, is expected to 
coincide with $\rho
(s)_{\rm exp}$, with the uncertainty of order $O[(\alpha_s (s))^3]$
and $O(1/s^3)$.
The uncertainty 
in the theoretical prediction of this order of magnitude is {\em natural} since
 terms of this order are neglected in
the theoretical calculation of  $D (Q^2)$. If the coincidence
in this corridor does take place,
we say that the quark-gluon prediction is dual to the hadronic
spectral density. If there are deviations going {\em beyond the 
natural uncertainty},
we call them violations of duality. 
Needless to say, once our calculation of 
$D (Q^2)$ becomes more precise, the definition of the
``natural uncertainty" in $\rho (s)_{\rm theor}$ changes accordingly.

This is the most clear-cut definition  
I can suggest. From the formal standpoint, it connects
the duality violation issue with that
of analytic continuation from the Euclidean to Minkowski domain.
Negligibly small corrections (legitimately) omitted in the
Euclidean calculations may and do get enhanced in Minkowski.
Exponentially small terms at positive $Q^2$ become oscillating and only power-suppressed
(at best) at positive $q^2$.

These oscillating terms defy the hierarchical ordering and can be referred to as 
``duality violating." The physical picture behind them is as follows.
The duality violations are  due to (i) rare  atypical events,
when the basic high-momentum quark transition occurs at large rather than short distances;
(ii) residual interactions occurring at large distances
between the quarks produced at short distances. In the first case
appropriate (Euclidean) correlation functions
develop singularities at finite $x^2$, while the second
mechanism is correlated with the $x^2\to\infty$ behavior.

In both cases the duality violating component follows the same 
pattern: {\em exponential  in Euclidean
and oscillating  in Minkowski}.  Three distinct regimes were
identified and considered in the literature so far:

\vspace{0.3cm}

$\bullet$ (i) Finite-distance singularities
\beq
s^{-\kappa /2}\, \sin (\sqrt{s} )\,;
\label{SS}
\eeq

\vspace{0.1cm}

$\bullet$ (ii) Infinite-distance singularities ($N_c = \infty$)
\beq
s^{-\eta /2}\, \sin (s)\,;
\label{SSS}
\eeq

\vspace{0.1cm}

$\bullet$ (iii) Infinite-distance singularities ($N_c $ large but finite, $s\to\infty$)
\beq
\exp{(-\alpha s)} \sin (s), \quad \alpha =
O\left(\frac{1}{N_c}\right) \ll 1\, .
\label{SSSS}
\eeq

\vspace{0.1cm}

These regimes are not mutually exclusive -- in concrete processes
one may expect
the duality violating component to be a combination of
 (i) and (ii), or (i) and (iii).
From the theoretical standpoint it is quite difficult
to consistently define the duality violating component of the type (3).
An operational definition I might suggest is as follows:
Start from the limit $N_c = \infty$ and identify
the component of the type (2).  Follow its evolution as 
$N_c$ becomes large but finite. 

Now a few words about global-versus-local dualities misconceptions
are in order. (This topic is also related to the issue of ordering of the limits
$N_c\to\infty$, $E$ fixed, or the other way around,
$E\to\infty$, $N_c$ fixed. These two limits are not interchangeable. As a result,
 $D(Q,N_c)$ must contain nonanalytic terms in $1/Q$ and $1/N_c$.)

Usually by local duality people mean
 point-by-point comparison of $\rho (s)_{\rm theor}$ and
$\rho (s)_{\rm exp}$, while global duality compares
the spectral densities  $\overline{\rho (s)}$ averaged
over some {\em ad hoc} interval of $s$, with an 
{\em ad hoc} weight function $w(s)$,
$$
\int_{s_1}^{s_2} ds\,  w(s)\,  \rho (s)_{\rm theor} \approx
\int_{s_1}^{s_2} ds \, w(s)\, \rho (s)_{\rm exp}\,.
$$
Many authors believe that  global duality defined in this way has a more solid
status than local duality. Some authors go so far as to say that
while global duality is certainly valid at high energies, this is
 not necessarily the case for local duality. This became a routine statement in the
literature. Well, routine does not mean correct.

In fact,
both procedures have exactly the same theoretical status.
The point-by-point comparison, as well as the comparison
of $\overline{\rho (s)}$'s (with an {\em ad hoc} weight function), must be
considered as distinct versions of local duality. 
The distinction between the ``local" quantities, such as   $R(e^+e^-)$ at a certain
value of
$s$ and the integrals of the type involved, say, in $R_\tau$  is 
quantitative rather
than qualitative. 
In both quantities there is a duality violating component,
the only distinction is in the concrete values of the indices of the power fall-off
in Eqs. (2) and (3), say,  3 vs. 6. Thus, the averaging 
over  $s$, as in $R_\tau$, makes the duality violation
somewhat more suppressed, but this is still something which we
totally miss in the {\em practical} version of OPE. In addition, at the moment these indices 
are model dependent. There is {\em no way} to reliably determine the value of duality violation, be it point 
by point as in $R(e^+e^-)$ or integrally, as in $R_\tau$,
from the analysis of the practical OPE {\em per se}.

The genuine global duality applies only to special integrals which can be 
{\em directly expressed through the
Euclidean quantities}. For instance, if the
integration interval extends from zero to infinity, and the weight function
is exponential, the integral
\beq
\int_{0}^{\infty} ds \exp\{-s/M^2\} \rho (s)\,,
\label{bortr}
\eeq
reduces  to the Borel transform of $D (Q^2)$ in the
 Euclidean domain (i.e. at positive $Q^2$). For such quantities,
duality can{\em not} be violated, by definition. \footnote{In mathematical literature
they refer to the transformation
(\ref{bortr}) as the inverse Laplace transform.
 It was first worked in 1930 by Post \cite{post}. Curiously, some questions that are difficult
 from the general mathematical standpoint
 seem to be rather trivial for theoretical physicists. I thank Martin Block and Arkady Vainshtein for this reference.
 See also \cite{bryan}.} 
   
   In the limit $E\to\infty$, $N_c$ fixed, dynamics {\em per se} ``globalizes"
   duality, since the resonance widths are switched on and produce a smearing of
   the spectral density. This dynamical smearing may increase the indices
   $\kappa$ and $\eta$ in (2), and (3), but in no way can eliminate
   deviations from duality in the sense I explained above. In general, the indices 
   $\kappa$ and $\eta$  are model-dependent. Why?
   
   By definition,  one 
 can{\em not} build  an exhaustive {\em theory} of the duality
violations based   on the
first principles. Indeed, assuming there is a certain dynamical mechanism
(which goes beyond perturbation theory and condensates)
for which such a theory exists, one would include
the corresponding component in the theoretical calculation.
The reference quantity, $D(Q^2)_{\rm theor}$, will be redefined
accordingly. After the analytic continuation to Minkowski,
this will lead, in turn,  to a new
theoretical spectral density to be used as a reference $\rho (s)_{\rm theor}$
in the duality relation. 

Thus, by the very nature of the problem, it is bound to be treated in models
of various degrees of fundamentality and reliability.
This is because the duality violation  parametrizes our ignorance.
Ideally, the models one should aim at
must have a clear-cut physical interpretation,
and must be tested,  in their key features, against experimental data.
This will guarantee a certain degree of confidence
when these models are  applied to the estimates
of the duality violations
in the processes and kinematical conditions where they had not
been  tested.

It is rather discouraging that there was very little progress, if at all, in this direction
in the last 10 years.

 \section{AdS/QCD vs. SVZ sum rules}
  
  It all started in 1998 when Maldacena; Gubser, Klebanov and Polyakov; and Witten
argued \cite{maldacena} that certain four-dimensional super-Yang--Mills theories
at large $N$ could be viewed as holographic images
of higher-dimensional string theory. In  the limit of a large 't Hooft coupling
the latter was shown to reduce to anti-de-Sitter supergravity.
The framework got the name ``Anti-de-Sitter/Conformal Field Theory (AdS/CFT) correspondence."

By now, it is generally believed that ten-dimensional string theory in suitable space-time backgrounds can have a dual, holographic description in terms of superconformal
gauge field theories in four dimensions.\footnote{Warning:
This gauge-gravity duality has nothing to do with the quark-hadron duality 
which was discussed 5 minutes ago.} Conceptually, the
 idea of a string-gauge duality ascends  to 't Hooft \cite{'tHooft:1973jz}, who realized that the perturbative expansion of  SU($N$) gauge field theories in the large-$N$  limit (with the 't Hooft coupling fixed)
 can be reinterpreted as a genus expansion of discretized two-dimensional surfaces built from the field theory Feynman diagrams. 
This  expansion resembles the string theory perturbative expansion  in the string coupling constant.
The AdS/CFT correspondence is a quantitative realization of this idea for four-dimensional gauge theories. In its purest form it identifies the ``fundamental type IIB superstring in a ten-dimensional anti-de-Sitter  space-time background AdS$_5\times S^5$
with the maximally supersymmetric ${\cal N} =4$ Yang--Mills theory with gauge group SU($N$) in four dimensions." The latter theory is superconformal.

Shortly after \cite{maldacena} a new ambitious goal was set:  to get rid
of conformality  and get as close to actual  QCD as possible.
There are two   lines of though.
Chronologically the first was the top-down approach 
pioneered by Witten; Polchinski and Strassler; Klebanov and Strassler;
Maldacena and Nu\~nez, and others. Here people try to obtain honest-to-god solutions
of the ten-dimensional equations of motion, often in the limit of
the large 't Hooft coupling when on the string side of the theory one deals with supergravity limit. The problem is: in many instances
these solutions
are dual to ... sort of QCD ... kind of QCD ..., rather than QCD as we know it. For instance, 
Witten's set-up or the Maldacena--Nu\~nez solution guarantee color confinement but 
the asymptotically free regime of QCD is not attained.

The Klebanov--Strassler supergravity 
solution  is near AdS$_5$ in the ultraviolet limit, a crucial property for the existence of a dual four-dimensional 
gauge theory. In the ultraviolet this theory exhibits logarithmic running of the couplings which goes under the name of the {\em duality cascade}. They start from string theory on
a warped deformed conifold and discover a cascade of
SU($kM)\times$SU($(k-1)M$) supersymmetric gauge theories on the other side.
As the theory flows to  the infrared, $k$ repeatedly changes by unity, 
see the review paper \cite{mast}. In the infrared
this theory exhibits a dynamical generation of the
scale parameter $\Lambda$, which manifests itself in the deformation of the conifold on the string side.

There is a variant  of the 
top-down approach in which the requirement of
the exact solution of the supergravity equations is
``minimally" relaxed. Confinement is enforced
through a crude cut-off 
of the AdS bulk in the infrared, at $z_0$, where $z$ is the fifth dimension.
 This leads to a ``wrong" confinement, however.
 In particular, the Regge trajectories do not come out
linear. A few years ago, preparing for a talk \cite{mishamig}, I suddenly realized
that the meson spectrum obtained in this way
identically coincides with the 30-year-old result \cite{AAM} of Alexander Migdal,
who, sure enough, had no thoughts of supergravity in five dimensions. 
His idea was to approximate logarithms of perturbation theory
by an infinite sum of poles in the ``best possible way."
Then this strategy was abandoned since it contradicts OPE.
Now, in essence, the Migdal program is   resurrected in a new incarnation
which goes under the name of AdS/QCD.
What was Migdal's goal? He asked himself:
``what is the best possible accuracy to which one-loop $\log Q^2$
can be approximated by an infinite sum of infinitely narrow
discrete resonances?" and ``What are the 
corresponding values of the resonance masses
and residues?"  He answered this question as follows:
``the accuracy is exponential at large $Q^2$
and the resonances must be situated at the zeros of a Bessel function."
This is exactly the position of the excited $\rho$ mesons
found in the first detailed AdS/QCD work~\cite{5}.

 The reason for the
coincidence of the 1977 and 2005 results is fully clear (both of them are admittedly wrong).
Abstractly speaking, one could have improved at least some aspects, for instance, if
instead of supergravity on the string side we could have kept the
 the full-blown string theory still adhering
to the above hard-wall approximation, we would   restore  asymptotic linearity of
the Regge trajectories at large angular momenta $J$ or excitation numbers $n$.
In this limit vein one could then calculate, in addition, say, the meson decay rates,
as was done recently by Sonnenschein and collaborators, who
recovered the 1979 Casher--Neuberger--Nussinov \cite{Casher:1978wy} quasiclassical formula!
However, nobody succeeded so far in
obtaining the full spectrum of crucial QCD regularities following this road.

This was the reason for the advent of AdS/QCD which I have just mentioned a couple minutes 
ago.
 I should add 
that the  bottom-up AdS/QCD guess-and-trial approach was pioneered by
Son and Stephanov \cite{SS}.
If in the  AdS/CFT correspondence (string-gauge duality) the five-dimensional metric $g_{AB}(x_\mu,z)$ is ``scientifically" {\em obtained from the Einstein equations} in five dimensions,
in AdS/QCD it is postulated {\em ad hoc}. From AdS/CFT 
we get, on ``our" side of duality, a
gauge theory in four dimensions which  is superconformal.  To get 
closer to QCD one must break SUSY, conformality and build in the property of asymptotic freedom.
  To this end
people abandon the five-dimensional  Einstein equations altogether
 and {\em try to guess} appropriate five-dimensional metric. A spectral quantum-mechanical
 equation with this conjectured metric gives us
the hadronic spectral densities. The guideline in the guessing process is the 
``marriage" between the holographic representation
and OPE-based methods,
plus chiral symmetry breaking, plus  other known regularities of the hadronic world
one can squeeze (with luck) in the construction.
The theorist is supposed to make various conjectures 
{\em en route}. The target is to build 
a  five-dimensional metric encoding as much information on real QCD
as possible, and then, with this metric in hands,
get new insights and make new predictions. 

It is clear that AdS/QCD and SVZ sum rule method share some common features.
It is no surprise that many features of the SVZ expansions were recovered.
The guessing of the five-dimensional metric is similar to 
developing the continuum model in the sum rules. Sum rule predictions for low-lying states are rather insensitive to the continuum model. This is an advantage. On the other hand, choosing the five-dimensional metric one completely specifies the relationship between the
low-lying states  and all higher excitations, with absolute sensitivity
to the entire spectrum. Some view this as an advantage, others as disadvantage.
For instance, the 
original Òhard cut-offÓ metric  \cite{5}
gives resonances at the zeros of the
Bessel function
(remember Migdal!). It gives
 parabolic Regge trajectories, and $\Pi(Q) \sim  \ln Q^2 + \exp(-Q)$ in Euclidean (remember Migdal!).
It was later replaced by a Òsoft cut-offÓ metric which gives equidistancy in $M^2$, linear Regge trajectories and ÒrigidÓ $1/Q^2$ corrections in the Euclidean expansion, as if all
condensates were expressed in terms of the lowest-order condensate.

By and large, I cannot say that at present AdS/QCD
gives a better (or more insightful) description
of the hadronic world, than the good old SVZ condensate-based method.
Given a rather crude character 
of the hard-wall and similar  approximations,
perhaps, today one may hope to  extract only  universal information 
on hadronic dynamics, steering clear of all details.  
However, I do not exclude that further studies of these two approaches,
in conjunction, will be beneficial for both and will add 
some significant understanding
to our knowledge of the hadronic world.
I appeal to young theorists involved in this area of research
to invest effort in this subject.

\section{Can holography  help?}

As it should be clear, I do not expect for the time being the AdS/QCD to produce 
detailed predictions
 superior (or even close) in their reliability and precision to those of the SVZ sum rules.
However,  I hasten to make a reservation. If one can find such general 
problems whose solution does not depend on  particular choices of the five-dimensional metric,
one has much better chances to obtain a successful 
prediction. Such an attempt has been undertaken  recently in \cite{gorsky,son} in a  
 problem
in a way related to the chiral anomaly.\footnote{That was probably
the original reason behind the belief that holography will work.}
As we will see, no breakthrough occurred, although the results obtained
in \cite{gorsky,son} give some food for further thought.

It is well known that the longitudinal part
of the fermion triangle graph is unambiguously and exactly fixed by the chiral anomaly.
In essence, this is a topological rather than dynamical quantity.
The knowledge of the above longitudinal part gives us the famous formula for
$\pi^0\to \gamma\gamma$. Vainshtein asked himself a question \cite{Vainshtein:2002nv}
(see also \cite{Czarnecki:2002nt,Knecht:2003xy}) whether 
the transverse part of the triangle graphs is also constrained.
He demonstrated that the transverse part $w_T(Q^2)$ of the current-current correlator in an
infinitesimally weak electromagnetic field  
defined as
\beqn
\label{eq:expansion}
\langle j_{\mu} j_{\nu}^5 \rangle_{\hat F} &=&-\frac{1}{4\pi^2}
\left[w_T(q^2)\big(-q^2 \tilde F_{\mu \nu} + q_{\mu}q^{\sigma} 
  \tilde F_{\sigma \nu}\right.
  \nonumber\\[3mm]
&-&\left.
q_{\nu}q^{\sigma} \tilde F_{\sigma \mu}\big) 
+ w_L(q^2)\, q_{\nu}q^{\sigma} \tilde F_{\sigma \mu} \right],
\eeqn
is {\em not} renormalized in perturbative QCD. However, because of the chiral symmetry breaking,
the above nonrenormalization theorem is not extendable to
to nonperturbative effects \cite{Vainshtein:2002nv}. Thus, $w_L$ and $w_T$
have different status: the latter quantity is dynamical. Nevertheless, the fact that
the $\alpha_s$ series is absent in $w_T$ gives one hope that
only some general aspects of QCD are involved in its calculation.
Under a simple additional assumption\ of the pion 
  dominance,
Vainshtein obtained the following analytic ``prediction"
for the
vacuum magnetic susceptibility $\chi$ introduced in \cite{Ioffe:1983ju}:
 \beq
\label{eq:chi}
\chi=-\frac{N_c}{4\pi^2 f_{\pi}^2}\,.
\label{7}
\eeq
Here $N_c$ is the number of colors, and $f_\pi\approx 92\,\,$MeV  is the pion constant.
I put ``prediction" in the quotation marks because there
was  no theoretical justification for the above-mentioned simplest assumption,
as was {\em certainly} noted and emphasized in the original paper \cite{Vainshtein:2002nv}.
I guess, the goal of the holography explorers 
in this issue was to find a justification for this or a similar relation, perfect the coefficients, and, in general,
elevate its status to the level where    the quotation marks could be dropped.

First, it was realized \cite {gorsky} that the gravity dual 
in the case at hand must be  Yang--Mills--Chern--Simons theory.
Addition of the Chern--Simons term turned out to be absolutely necessary.
In the so-called hard-wall version of holography the vacuum magnetic susceptibility
was found (numerically) \cite{gorsky} to be close to (\ref{7}), with the coefficient
larger than $N_c/4\pi^2$ by about 10\%.

Then other versions of holography, such as the 
so-called soft model of  the ``bottom-up" AdS/QCD \cite{SS}
 and the ``top-down" Sakai--Sugimoto model
\cite{Sakai:2004cn} (both are popular in this range of questions)
were explored in \cite{son}. 
In this rather broad  class of Yang--Mills--Chern--Simons dual theories,
with the chiral symmetry broken by the boundary conditions in the infrared,
independently of the choice of the gravity metrics,
the following relation takes place \cite{son}:
\beq
\label{eq:trans}
w_T(Q^2)=\frac{N_c}{Q^2}-\frac{N_c}{f_{\pi}^2}
  \left[\Pi_{A}(Q^2) - \Pi_{V}(Q^2)\right]_F,
  \label{6}
\eeq
for any $Q^2$. Here $\Pi_{A,V}$ are the two-point functions of the axial-vector (vector)
currents in the background (very soft)
electromagnetic field. 
The first term in (\ref{6})  was obtained by Vainshtein.
The second term obviously vanishes to any finite order in perturbation theory.
This is a nonperturbative correction found through holography but independent of the
particulars of the five-dimensional metric. Equation (\ref{6}) implies a new
set of  relations for various resonance parameters. 

Considering Eq. (\ref{6}) at {\em large} $Q^2$, using the SVZ-type operator product expansion for
$ \left[\Pi_{A}(Q^2) - \Pi_{V}(Q^2)\right]_F$,
and factorization for the four-quark matrix element (justified by the large-$N_c$ limit)
one can derive  from (\ref{6})  a consistency condition \cite{son} for the
vacuum magnetic susceptibility, in an analytic form.
Remarkably, this is exactly the same formula (\ref{7}) obtained  by Vainshtein.

 Another example of predictions \cite{son} for physical 
observables following from (\ref{6}) are the sum rules
\beqn
\label{eq:sum-rule1}
\sum_{j}\frac{{g_{\gamma V_i A_j}}g_{A_j}}{m_{A_j}^2-m_{V_i}^2}
&=&
-\frac{N_c}{4\pi^2 f_{\pi}^2}g_{V_i},
\\[4mm]
\label{eq:sum-rule2}
\sum_{i}\frac{{g_{\gamma V_i A_j}}g_{V_i}}{m_{A_j}^2-m_{V_i}^2}
&=&
-\frac{N_c}{4\pi^2 f_{\pi}^2}g_{A_j}\,.
\eeqn
In the first sum rule $i$ is fixed (and arbitrary) while
$j$ runs over all axial-vector resonances. Likewise,
in the second expression $j$ is fixed while $i$
runs over all vector resonances. For broad resonances one can substitute the sums by the integrals
in the spirit of the SVZ method.

Let us ask ourselves how successful numerically is the Vainshtein formula.
The pion constant in (\ref{7}) is normalized as $f_\pi \approx 92\,$MeV. Substituting this number we
arrive at $\chi \approx - 9.0$ GeV$^{-2}$. At the same time, the magnetic susceptibility
had been determined from the QCD sum rules long ago.
Unfortunately,  calculation of the magnetic susceptibility, presented in the
main text of \cite{Ioffe:1983ju} is not quite correct since an inappropriate 
dispersion relation was used.  The  correct result is quoted  in `Note added in proof'  in the same paper.
The most precise  evaluation of the magnetic susceptibility can be found in Sec. 6.3 
of the book \cite{IoffeBook},
\beq
\chi_{\rm QCD\,\, SR} =
-3.15 \pm 0.3 \,\,{\rm GeV}^{-2}
\label{11}
\eeq
for the normalization point around 1 GeV. The discrepancy is about a factor of 3, somewhat larger than
could have been anticipated.

The main problem with holography 
which clearly reveals itself in confronting (\ref{7}) and (\ref{11})
is that holography, as we know it now,  does not reproduce those several terms of OPE
which are solidly established. Therefore,  fine details of the hadronic picture 
obtained from holography come out  wrong (at least, for the
time being), although some overall contours are, perhaps, captured right.

 \section{Instead of Conclusions}

The richness of the hadronic world is enormous.

 It describes a very wide range of natural phenomena, e.g.:

$\bullet$ all of nuclear physics;

$\bullet$ Regge behavior and Regge trajectories (highly excited meson and baryon states);

$\bullet$ strongly coupled quark-gluon plasma; high-$T$  phenomena; 
color superconductivity at  high density (through color-flavor locking); neutron stars;

$\bullet$ richness of the hadronic world (chiral phenomena,
light and heavy quarkonia, the Zweig rule, glueballs and exotics,
exclusive and inclusive processes);

$\bullet$ hadronization of fast moving  colored sources, i.e. jets (of special interest 
are, of course, nonperturbative aspects of the jet physics);

$\bullet$ interplay between strong and 
weak interactions (in particular, the so-called penguin mechanism);

\vspace{1mm}
\noindent
 and many other issues.
 
 \vspace{1mm}

At short distances QCD is weakly coupled, allowing high precision 
perturbative (multi-loop, multi-leg) calculations.
However, the advent of the era of 
arbitrarily exact analytical computations at all energies and momenta, especially
in the Minkowski domain, is not expected in the foreseeable future,
 due to strong coupling nature  of the large-distance dynamics.
Let us ask ourselves: what do we want from this theory? Is it reasonable to
expect  high-precision analytic predictions  for all low-energy observables? Can we (and should we) compute  hadronic masses, matrix elements or, say,
proton's magnetic moment up to five digits? 

Unlike QED, most probably  we will never be able to analytically calculate the above-mentioned 
and similar
observables
with this precision. But do we really need this?
To my mind, what is really needed is the completion
of the overall qualitative picture of confinement 
 + development of various
 reliable  approximate techniques custom-designed for specific applications.
 The original sum rule method, extended by numerous later developments,
 fits very well.
In this context
QCD sum rules do have a future, don't they?

\section*{Note Added in December}

A very recent publication \cite{komar} admired me by its elegance.
Zohar Komargodski implemented, in a brilliant way,
the old idea 
\cite{olid} that all vector mesons of QCD (i.e. $\rho$ and its excitations)
are in fact Higgsed gauge bosons of a hidden non-Abelian local symmetry (or symmetries)
of the hadronic world. 
This was done in a fully controllable setting of supersymmetric QCD and is heavily based on 
Seiberg's duality \cite{nsei}. The dream of Yang and Mills, who originally designed
the Yang--MIlls theory in the context of the description of the hadronic world, is realized!
The same idea \cite{olid} that served as an important catalyst  in the advent of AdS/QCD \cite{SS},
in its supersymmetric reincarnation 
provided an analytic and parametric  proof of the vector meson dominance,
a phenomenon that is a crucial feature in the QCD sum rules (albeit seen there numerically rather than parametrically). This clearly tells us that the tool kit available to us
for dealing with mysteries of QCD is expanding, and
still unsolved mysteries still have chances to be solved in the future.

\section*{Acknowledgments}
I want to say thank you to S. Narison who kindly invited me to deliver this minilecture.
I am grateful to Arkady Vainshtein for countless useful discussions of all issues
covered in this talk. Useful comments from B. Ioffe and D. Son are acknowledged.
This write-up was done during my sabbatical at 
the Jefferson Physical Laboratory, Harvard University. I would like to thank my 
Harvard colleagues for hospitality.

This work is supported in part by the  DOE grant DE-FG02-94ER-40823.


\begin{thebibliography} {999}

\bibitem{unpubintro}
M.~Shifman,
{\em Vacuum structure and QCD sum rules: Introduction,}
  Int.\ J.\ Mod.\ Phys.\  A {\bf 25},  226  (2010).
  See also an extensive list of references therein.
  
\bibitem{persistent}
M.~Shifman,
{\em Persistent challenges of quantum chromodynamics,}
  Int.\ J.\ Mod.\ Phys.\  A {\bf 21}, 5695 (2006)
  [arXiv:hep-th/0606015].

\bibitem{SW}
N.~Seiberg and E.~Witten,
 {\em Electric-magnetic duality, monopole
 condensation, and confinement in ${\mathcal N} = 2$ supersymmetric Yang--Mills theory,}
Nucl. Phys. {\bf B426}, 19 (1994),
(E) {\bf B430},  485 (1994) [hep-th/9407087];
 {\em Monopoles, duality and chiral symmetry breaking in ${\mathcal N} = 2$
 supersymmetric QCD,}
Nucl. Phys. {\bf B431}, 484  (1994)
[hep-th/9408099].

\bibitem{Wilson}
K.~G.~Wilson,
{\em Nonlagrangian models of current algebra,}
  Phys.\ Rev.\  {\bf 179}, 1499 (1969);
K.~G.~Wilson and J.~B.~Kogut,
{\em The Renormalization group and the epsilon expansion,}
  Phys.\ Rept.\  {\bf 12}, 75 (1974).
  
\bibitem{Belavin:1984vu}
  A.~A.~Belavin, A.~M.~Polyakov and A.~B.~Zamolodchikov,
{\em Infinite conformal symmetry in two-dimensional quantum field theory,}
  Nucl.\ Phys.\  B {\bf 241}, 333 (1984).
  
\bibitem{NSVZ2}
V.~A.~Novikov, M.~A.~Shifman, A.~I.~Vainshtein and V.~I.~Zakharov,
 {\em Wilson's Operator Expansion: Can It Fail?}
  Nucl.\ Phys.\  B {\bf 249}, 445 (1985)
  
\bibitem{snap}
M.~A.~Shifman,
 {\em Snapshots of hadrons or the story of how the vacuum medium determines  the
 properties of the classical mesons which are produced, live and die  in the
 QCD vacuum,}
  Prog.\ Theor.\ Phys.\ Suppl.\  {\bf 131}, 1 (1998)
  [arXiv:hep-ph/9802214].
  
  \bibitem{nsvzsigma}
V.~A.~Novikov, M.~A.~Shifman, A.~I.~Vainshtein and V.~I.~Zakharov,
{\em Two-Dimensional Sigma Models: Modeling Nonperturbative Effects Of Quantum
  Chromodynamics,}
  Phys.\ Rept.\  {\bf 116}, 103 (1984).
  
  \bibitem{Shif1} 
M. Shifman, 
{\em Theory of Preasymptotic Effects in Weak Inclusive Decays},
in Proc. Workshop on {\em Continuous Advances in QCD}, ed. A. 
Smilga
(World Scientific, Singapore, 1994), page 249 [hep-ph/9405246];
{\em Recent Progress in the Heavy Quark Theory}, in {\it  Particles, 
Strings and Cosmology},  Proc.
of the XIX
Johns Hopkins Workshop on Current Problems in Particle Theory
and the V PASCOS Interdisiplinary Symposium, Baltimore,
   March 1995, 
Ed. 
   J. Bagger (World Scientific, Singapore, 1996), page 69
[hep-ph/9505289].

\bibitem{QHD}
M.~Shifman,
{\em Quark-hadron duality,} 
in {\sl At the Frontiers of Particle Physics: Handbook of QCD}, 
Ed. M. Shifman (World Scientific, Singapore, 2001), Vol. 3, p. 1447,
  arXiv:hep-ph/0009131.
  
 \bibitem{alike}
  V.~A.~Novikov, M.~A.~Shifman, A.~I.~Vainshtein and V.~I.~Zakharov,
{\em Are All Hadrons Alike?},
  Nucl.\ Phys.\  B {\bf 191}, 301 (1981).
  
\bibitem{svregge}
M.~Shifman and A.~Vainshtein,
{\em Highly Excited Mesons, Linear Regge Trajectories and the Pattern of the
  Chiral Symmetry Realization,}
  Phys.\ Rev.\  D {\bf 77}, 034002 (2008)
  [arXiv:0710.0863 [hep-ph]].
  
     \bibitem{post}
E.  Post, {\em Generalized Differentiation}, Trans. Amer. Math. Soc. {\bf 32}, 723--781 (1930).

   \bibitem{bryan}
   Kurt Bryan, {\em
Elementary Inversion of the Laplace Transform},
Rose-Hulman Institute of Technology Report MS-TR 98-07 (1999),
http://www.rose-hulman.edu/~bryan/invlap.pdf

\bibitem{maldacena}
J.~M.~Maldacena,
  Adv.\ Theor.\ Math.\ Phys.\  {\bf 2}, 231 (1998)
  [Int.\ J.\ Theor.\ Phys.\  {\bf 38}, 1113 (1999)]
  [arXiv:hep-th/9711200];
S.~S.~Gubser, I.~R.~Klebanov and A.~M.~Polyakov,
  Phys.\ Lett.\  B {\bf 428}, 105 (1998)
  [arXiv:hep-th/9802109];
E.~Witten,
  Adv.\ Theor.\ Math.\ Phys.\  {\bf 2}, 253 (1998)
  [arXiv:hep-th/9802150].
  
\bibitem{'tHooft:1973jz}
  G.~'t Hooft,
{\em A Planar Diagram Theory For Strong Interactions},
  Nucl.\ Phys.\  B {\bf 72}, 461 (1974).
  
  \bibitem{mast}
  M.~J.~Strassler,
{\em The duality cascade,}  hep-th/0505153, [in {\sl Progress in String Theory: 
TASI 2003 Lecture Notes}, Ed. J. Maldacena (World Scientific, Singapore, 2005), pp. 419-510].
  
  \bibitem{mishamig}
  M.~Shifman,
{\em Highly excited hadrons in QCD and beyond,}
in {\em Quark-Hadron Duality and the Transition to pQCD}, Eds. A. Fantoni, S. Liuti, 
and O. Rond\'on (World Scientific, Singapore, 2006), pp. 171-192,
  arXiv:hep-ph/0507246.
  
  \bibitem{AAM}
 A.~A.~Migdal,
  {\em Series Expansion For Mesonic Masses In Multicolor QCD,}
  Annals Phys.\  {\bf 110}, 46 (1978);
H.~G.~Dosch, J.~Kripfganz and M.~G.~Schmidt,
{\em Structure Of Hadron In Migdal's Regularization Scheme For QCD,}
  Phys.\ Lett.\ B {\bf 70}, 337 (1977).
  
  \bibitem{5}
J.~Erlich, E.~Katz, D.~T.~Son and M.~A.~Stephanov,
{\em QCD and a holographic model of hadrons}, 
 Phys.\ Rev.\ Lett.\  {\bf 95}, 261602 (2005)
  [arXiv:hep-ph/0501128];
  see also A.~Karch, E.~Katz, D.~T.~Son and M.~A.~Stephanov,
{\em Linear Confinement and AdS/QCD,}
  Phys.\ Rev.\  D {\bf 74}, 015005 (2006)
  [arXiv:hep-ph/0602229].
  
\bibitem{Casher:1978wy}
  A.~Casher, H.~Neuberger and S.~Nussinov,
{\em Chromoelectric Flux Tube Model Of Particle Production,}
  Phys.\ Rev.\  D {\bf 20}, 179 (1979).
 
    \bibitem{SS}
    D.~T.~Son and M.~A.~Stephanov,
{\em QCD and dimensional deconstruction,}
  Phys.\ Rev.\  D {\bf 69}, 065020 (2004)
  [arXiv:hep-ph/0304182].
  
   \bibitem{gorsky}
   A.~Gorsky and A.~Krikun,
{\em Magnetic susceptibility of the quark condensate via holography,}
  Phys.\ Rev.\  D {\bf 79}, 086015 (2009)
  [arXiv:0902.1832 [hep-ph]].
   
  \bibitem{son} 
  D. Son and  N.Yamamoto, 
{\em Holography and Anomaly Matching for Resonances},
arXiv:1010.0718.

 \bibitem{Vainshtein:2002nv}
  A.~Vainshtein,
{\em Perturbative and nonperturbative renormalization of anomalous quark
triangles,}
  Phys.\ Lett.\  B {\bf 569}, 187 (2003)
  [arXiv:hep-ph/0212231].
  
  \bibitem{Czarnecki:2002nt}
  A.~Czarnecki, W.~J.~Marciano and A.~Vainshtein,
{\em Refinements in electroweak contributions to the muon anomalous magnetic
 moment,}
  Phys.\ Rev.\  D {\bf 67}, 073006 (2003)
  [Erratum-ibid.\  D {\bf 73}, 119901 (2006)]
  [arXiv:hep-ph/0212229].
  
  \bibitem{Knecht:2003xy}
  M.~Knecht, S.~Peris, M.~Perrottet and E.~de Rafael,
{\em New nonrenormalization theorems for anomalous three point functions,}
  JHEP {\bf 0403}, 035 (2004)
  [arXiv:hep-ph/0311100].
  
    \bibitem{Ioffe:1983ju}
  B.~L.~Ioffe and A.~V.~Smilga,
{\em Nucleon Magnetic Moments And Magnetic Properties Of Vacuum In QCD,}
  Nucl.\ Phys.\  B {\bf 232}, 109 (1984).
  
    \bibitem{Sakai:2004cn}
  T.~Sakai and S.~Sugimoto,
{\em Low energy hadron physics in holographic QCD,}
  Prog.\ Theor.\ Phys.\  {\bf 113}, 843 (2005)
  [arXiv:hep-th/0412141];
{\em More on a holographic dual of QCD,}
  Prog.\ Theor.\ Phys.\  {\bf 114}, 1083 (2005)
  [arXiv:hep-th/0507073].
  
   \bibitem{IoffeBook}
  B. L. Ioffe, V. S. Fadin, and L. N. Lipatov,
  {\sl Quantum Chromodynamics: Perturbative and Nonperturbative Aspects},
  (Cambridge University Press, 2010).
  
\bibitem{komar}
 Z.~Komargodski,
{\em Vector Mesons and an Interpretation of Seiberg Duality,}
  arXiv:1010.4105 [hep-th].
  
  \bibitem{olid}
  M.~Bando, T.~Kugo, S.~Uehara, K.~Yamawaki and T.~Yanagida,
 {\em Is Rho Meson A Dynamical Gauge Boson Of Hidden Local Symmetry?,}
  Phys.\ Rev.\ Lett.\  {\bf 54}, 1215 (1985);
  M.~Bando, T.~Kugo and K.~Yamawaki,
{\em Nonlinear Realization and Hidden Local Symmetries,}
  Phys.\ Rept.\  {\bf 164}, 217 (1988).
  
  \bibitem{nsei}
   N.~Seiberg,
 {\em Electric - magnetic duality in supersymmetric nonAbelian gauge theories,}
  Nucl.\ Phys.\  B {\bf 435}, 129 (1995)
  [arXiv:hep-th/9411149].

\end{thebibliography}
\end{document}